# Focal Loss based Residual Convolutional Neural Network for Speech Emotion Recognition

Suraj Tripathi[1], Abhay Kumar[1]*, Abhiram Ramesh[1]*, Chirag Singh[1]*, Promod Yenigalla[1]

[1]Samsung R&D Institute India – Bangalore
`suraj.tri@samsung.com, abhay1.kumar@samsung.com,`
`a.ramesh@samsung.com c.singh@samsung.com, promod.y@samsung.com`

**Abstract.** This paper proposes a Residual Convolutional Neural Network (ResNet) based on speech features and trained under Focal Loss to recognize emotion in speech. Speech features such as Spectrogram and Mel-frequency Cepstral Coefficients (MFCCs) have shown the ability to characterize emotion better than just plain text. Further Focal Loss, first used in One-Stage Object Detectors, has shown the ability to focus the training process more towards hard-examples and down-weight the loss assigned to well-classified examples, thus preventing the model from being overwhelmed by easily classifiable examples. After experimenting with several Deep Neural Network (DNN) architectures, we propose a ResNet, which takes in Spectrogram or MFCC as input and supervised by Focal Loss, ideal for speech inputs where there exists a large class imbalance. Maintaining continuity with previous work in this area, we have used the University of Southern California's Interactive Emotional Motion Capture (USC-IEMOCAP) database's Improvised Topics in this work. This dataset is ideal for our work, as there exists a significant class imbalance among the various emotions. Our best model achieved a 3.4% improvement in overall accuracy and a 2.8% improvement in class accuracy when compared to existing state-of-the-art methods.

## 1 INTRODUCTION

Standard Natural Language Processing (NLP) systems first transcribe speech into text and then apply deep learning techniques to recognize emotion in speech. Converting speech into text provides us with contextual data required for emotion recognition, but this conversion makes the system speaker independent and deprives the network of valuable spectral information that could be key to characterizing emotion in speech. Speech features, such as Spectrogram and MFCC, provide variations of speech over frequency and time ideal for Speech Emotion Recognition (SER) systems, which are dependent on the speaker's speech parameters such as pitch, amplitude, etc.

---
* equal contribution

In recent years CNN, which turns speech signals into feature maps, has been widely used in such areas of research [1] [2]. The drawback here is that such CNN based SER systems do not extend beyond a few convolutional layers as convergence is slow and the models become difficult to train. Typical architectures consist firstly of some convolutional layers, then a few recurrent layers and finally ending with fully connected feedforward layers. Such networks are not deep enough to extract rich information to be able to classify the input among different emotion classes accurately. ResNets [3] on the other hand allow us to have a very deep architectures ideal for obtaining very deep features.

Taking the many limitations of the traditional CNN based SER systems into consideration; we propose a speech features based 18-layer ResNet architecture, supervised by Focal Loss, which outperforms the state-of-the-art SER accuracies. Focal Loss was chosen over the widely used Softmax Loss as the latter has a tendency to be influenced heavily by trivial examples in case of data imbalance [4]. A big proportion of spoken utterances in day-to-day life have no strong emotion associated with it, i.e. they are Neutral in nature. Therefore, training under such an environment with large class imbalance results in a Softmax based system focusing itself more towards such easy examples (Neutral), whereas it should be giving more priority towards the harder emotion classes. Focal Loss, which is a dynamically scaled version of Softmax Loss, focuses the training towards the sparse set of hard examples while preventing the influence of a large number of easy examples on the system.

The main contributions of the current work are:

- The first attempt, as per our knowledge, to use Focal Loss in addressing the issue of class imbalance in Speech Emotion Recognition (SER) systems
- Proposed speech features based ResNet architecture, trained on IEMOCAP data, which outperforms the state-of-the-art emotion recognition accuracies

## 2 RELATED WORK

Deep Learning techniques have been the reason for significant breakthroughs in Natural Language Understanding (NLU) in the last few years. Baseline models, which don't work on the principle of deep learning, were significantly improved by Deep Belief Networks (DBN) for SER, proposed by Kim *et al.* [5] and Zheng *et al.* [6]. [1] used Spectrograms with Deep CNNs whereas Fayek *et al.* [7] made use of deep hierarchical architectures, data augmentation, and regularization with a DNN for SER. Ranganathan *et al.* [8] experimented with Convolutional Deep Belief Networks (CDBN), which learn salient multimodal features of expressions, to achieve good accuracies. Satt *et al.* [2] used deep CNNs in combination with Long Short Term Memory (LSTM) cells to achieve better results on the IEMOCAP dataset.

In recent years Recurrent Neural Networks (RNNs), capable of modelling long histories, have been used extensively in sequential modeling. LSTMs and Bidirectional LSTMs, which fix the gradient vanishing problem, have also been used in ASR sys-tems. Graves *et al.* [9] were the first to apply bidirectional training to LSTM networks to classify phonemes frame wise in continuous speech recognition. Lee *et al.* [10] used a bi-directional LSTM model to train feature sequences and achieved an emotion recognition accuracy of 62.8% on the IEMOCAP [11] dataset. LSTMs, however, become a computation bottleneck for very large sequences as they, at each time stamp, store multiple neural gate responses. CNNs were introduced to ASR as an alternative to LSTMs as they were much easier on computing power. Abdel-Hamid *et al.* [12], Amodei *et al.* [13], and Palaz *et al.* [14] were one of the earliest to use CNNs in ASR, but they only employed a few convolutional layers. Qian *et al.* [15] used very deep CNNs for effective recognition of noisy speech.

As very deep CNNs were observed to suffer from a slow rate of convergence and performance saturation and degradation, ResNets proved to be a good alternative with its skip connections in residual blocks. Luo *et al.* [16] considered the high performance of Recurrent Neural Networks and ResNets in speech and image related classification tasks to propose an ensemble SER system, which outperformed the best single-classifier, based SER system. Tzirakis *et al.* [17] utilized a CNN and a 50-layered ResNet to extract features from speech and visual data respectively.

Loss functions that deal with class imbalance have been a topic of interest in recent times. Lin *et al.* [4] proposed a new loss called Focal loss, which addresses class imbalance by dynamically scaling the standard cross-entropy loss such that the loss associated with easily classifiable examples are down-weighted. They used it in the field of Dense Object Detection and were able to match the speed of previous one-stage detectors while surpassing the accuracy of all existing state-of-the-art two-stage detectors. Yang *et al.* [18] proposed skip-connections in CNN structures trained under Focal loss to enhance feature learning for Vehicle Detection in Aerial Images.

## 3 PROPOSED METHODS

Deep learning methods have been successfully applied on speech features to extract high-order non-linear relationships. CNNs, in particular, have been used extensively to gather information from raw signals in various applications such as speech recognition, image recognition, etc. [19] [20]. Deep CNNs improve generalization and easily outperform shallow networks, but they have a tendency to converge slowly and can be difficult to train. ResNets were proposed to ease this difficulty in training very deep CNNs. ResNets, when used with speech features such as Spectrogram and MFCC, provide the required high-level features to better capture emotion in speech. Experiments have been performed on speech features supervised by Focal Loss to address the class imbalance in IEMOCAP dataset and achieve high accuracies.

### 3.1 Model Architecture

When a network becomes very deep, we encounter a couple of unavoidable problems. One is that the gradients tend to either vanish or explode. The other is that as depth increases the accuracies tend to either saturate or fall. ResNets, first introduced by He *et al.* [3], consists of a number of stacked residual blocks with outputs from the lower layers linked to the inputs of the higher layers. These "shortcut connections" turn the input maps into identity maps.

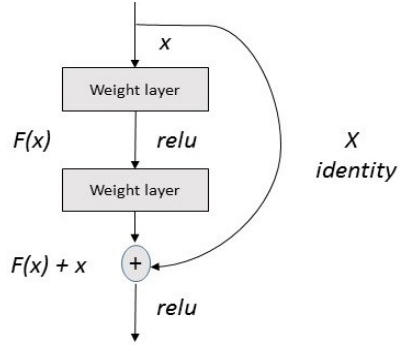

**Fig. 1.** Residual block

The residual block described in Figure 1 is defined as:

$$y = F(x, W_i) + x \qquad (1)$$

Where $x$ and $y$ are the input and the output layers respectively, and $F(x, W_i)$ is the stacked non-linear mapping function. These residual connections help improve convergence speed during training and do not degrade in performance with an increase in depth.

The residual function $F(x, W_i)$ is flexible and could contain multiple layers, 2 in our case. In addition, the above notations are applicable to multiple convolutional layers and not just to fully connected layers. Our proposed network, shown in Figure 2, is an 18-layer ResNet, which takes speech features as input. We apply Batch Normalization before computing the Rectified Linear Unit (ReLU) activations. Contrary to the norm, we have introduced a Focal loss module after the last hidden layer.

### 3.2  Feature Extraction

The presented models use Spectrogram and MFCC as input to the ResNet. As the audio files in the IEMOCAP's Improvised Topic dataset vary in duration, the length of each clip was restricted to 6 seconds or less. This was done under the assumption that the feature variations, which could possibly characterize the emotion in a speech signal, will be present throughout the dialogue, and hence will not be lost by this reduction in audio length. The Mel-frequency scale, which the Spectrogram and MFCC are scaled to, puts emphasis on the lower end of the frequency spectrum over the higher ones, thus imitating the perceptual hearing capabilities of humans. We used the "librosa" python package to compute the Spectrogram and Cepstral coefficients.

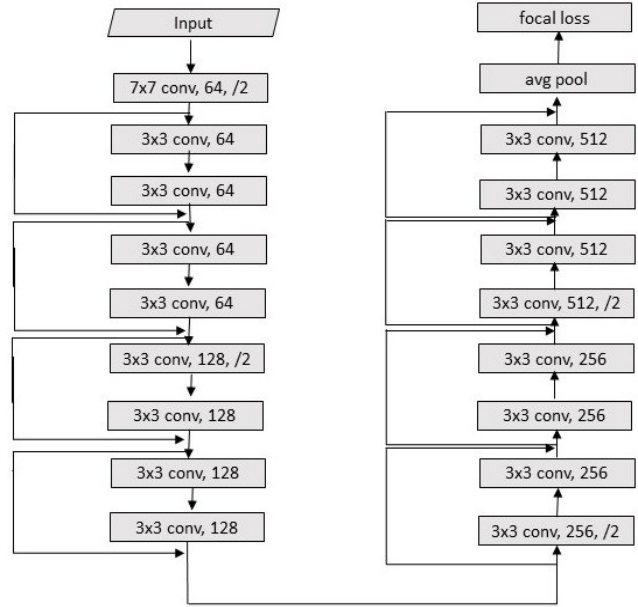

**Fig. 2.** Model architecture

**Spectrograms**

Mel-Spectrogram is a 2D representation of log-magnitude intensity (dB) over frequency and time. The audio signal is sampled at the sampling rate of 22050Hz. Subsequently, each audio frame is windowed using a "*hann*" window of length 2048 to increase its continuity at the endpoints. To calculate the power spectrum of each frame, we apply Short Term Fourier Transform (STFT) on windowed audio samples. We use Fast Fourier Transform (FFT) windows of length 2048 and an STFT hop-length equal to 512.

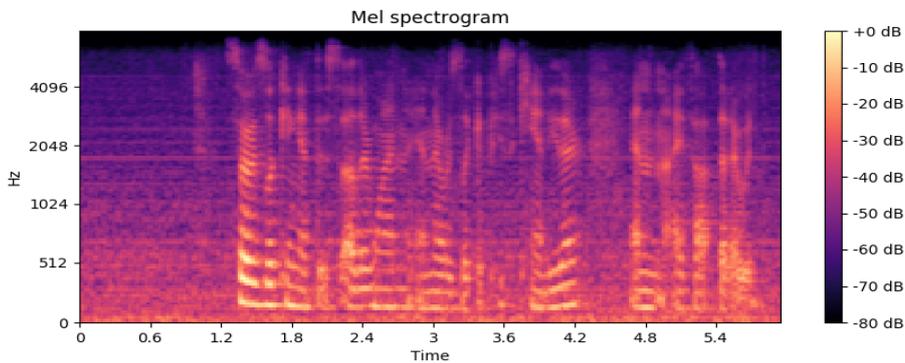

**Fig. 3.** Sample spectrogram for an audio recording from IEMOCAP dataset.

The obtained Spectrogram magnitudes are then converted to Mel-scale to get Mel-frequency spectrum. 128 Spectrogram coefficients per window are used in our model.

**MFCC**

Mel-frequency Cepstrum (MFC) is a 2D representation of the Short-Term Power Spectrum of sound. It is based on a linear cosine transform of a log power spectrum on a non-linear Mel-scale of frequency. The parameters used in the generation of MFCC are same as the ones described for Spectrogram. The only additional step for MFCC generation compared to Spectrogram is that a Discrete Cosine Transform (DCT) is performed on the obtained coefficients. 40 MFCCs per window were generated compared to the 128 for Spectrogram.

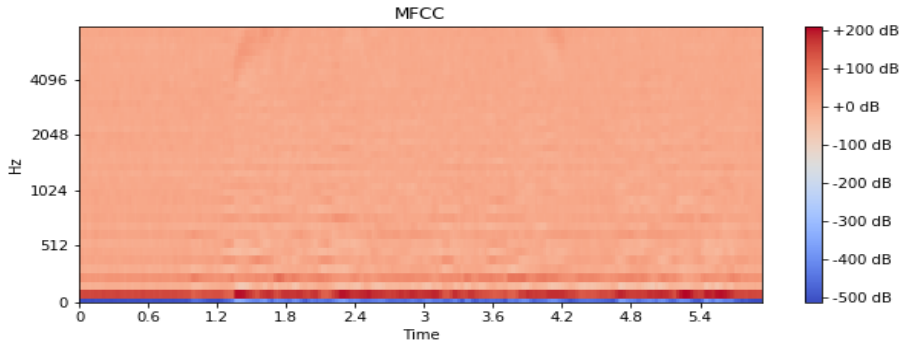

**Fig. 4.** Sample MFCC for an audio recording from IEMOCAP dataset.

### 3.3 Focal Loss

A property of the widely used Cross-Entropy Loss is that even easily classifiable examples ($p_t \geq 0.5$) result in a loss with a significant magnitude [4]. The losses incurred during training from easy examples, which constitute the majority of the dataset, can have a negative impact on the rarer classes. The Neutral emotion, which covers a majority of the dataset, tends to comprise the bulk of the loss and ends up dominating the gradient. Focal Loss maneuvers this by reshaping the Cross-Entropy Loss function by giving less importance to the easy examples and focusing more on the hard ones. A general way of formulating Focal loss is:

$$FL(p_t) = -(1 - p_t)^\gamma \log(p_t) \qquad (2)$$

Where $p_t$ is the model's estimated probability for the class, and $\gamma \geq 0$ is the tunable focusing parameter.

For inputs that are misclassified their contribution to total loss is not affected as the modulating factor is close to one and $p_t$ is small. For inputs that are classified correctly, the modulating factor becomes very small, and the loss is significantly down-weighted thus reducing its contribution to total loss even if they are large in number and preventing the model from being overwhelmed by such examples.

For instance when $p_t$ = 0.8 and $\gamma$ = 2 Focal Loss is 96% lower compared to Cross-Entropy Loss, which means that the correctly classified examples' effect on the gradients is reduced heavily. When $p_t$ is already low (< 0.1) the loss pretty much remains the same aiding the model in learning these hard examples.

## 4   Dataset

The University of Southern California's Interactive Emotional Motion Capture (USC-IEMOCAP) database consists of five sessions, each session comprising of a conversation between two people, in both scripted and improvised topics. Gender bias is minimized as each session is acted upon and voiced by both male and female voices. The data collected is first split into utterances of length varying between 3-15 seconds and then labelled by 3-4 evaluators. To label the emotion, the assessors had to choose among 10 different emotion classes (Happiness, Surprise, Fear, Sadness, Frustration, Excited, Anger, Disgust, Neutral and other). We limited our emotion classes to just four (Anger, Happiness, Sadness and Neutral) to remain consistent with earlier research. We chose only those utterances where at least 2 experts were in agreement with their decided labels and only used Improvised Topic data, again being consistent with prior research. We excluded scripted data as it showed too strong a correlation with labeled emotions, which could lead to lingual content learning. The proportions of the four classes in the final experimental dataset are Neutral (48.8% of the total dataset), Happiness (12.3%), Sadness (26.9%) and Anger (12%). As there exists an imbalance for data between different emotional classes, we present our results on overall accuracy; average class accuracy and also show the confusion matrix (refer Table 2 and 3).

## 5   EVALUATION AND DISCUSSION

We have shown the effectiveness of the proposed methods for emotion detection with our benchmark results on IEMOCAP dataset and compared with the previous related research. We have split the dataset into training and test sets using stratified K-folds. Comparison of our 5-fold cross-validation experimental results is made with some of the recent results on emotion classification and is presented in Table 1. The proposed ResNet models are based on Spectrogram and MFCC inputs and are supervised by Focal Loss, with our best model achieving an overall accuracy improvement of 3.4% and a class accuracy improvement of 2.8%. Overall accuracy is a measure of total counts irrespective of class, and class accuracy is the mean of accuracies obtained in each of the emotion classes.

Table 1. Comparison of accuracies

| Methods | Input | Overall Accuracy | Class Accuracy |
|---|---|---|---|
| Lee *et al.* [10] | Spectrogram | 62.8 | 63.9 |
| Satt *et al.* [2] | Spectrogram | 68.8 | 59.4 |
| Yenigalla [21] | Spectrogram | 71.2 | 61.9 |
| Proposed Model | Spectrogram | **74.2** | **64.3** |
| Proposed Model | MFCC | **74.6** | **66.7** |

## 5.1 Ablation study of the effectiveness of focal loss:

As mentioned earlier, the dataset is not well balanced, with Neutral constituting almost half of it. Tables 2 and 3 represent the confusion matrix showing misclassification rates between each pair of classes for the MFCC based ResNet, but independently trained on Softmax Loss or Focal Loss, for an equal number of epochs, respectively. Table 2 clearly validates the problems with using Softmax Loss in situations where there is a significant class imbalance in the dataset. The network has tuned itself to work very well in classifying the most abundant emotion, Neutral, with an 89.1% accuracy, but suffers in classifying the rarer emotions correctly. However, when the network is trained on Focal Loss, we can clearly see in Table 3 the improvement in recognition accuracies for the rarer classes. The rate of recognition for Happiness almost doubles; Sadness and Anger also observe significant improvement but with a slight drop in Neutral.

Table 2. Confusion matrix (ResNet on MFCC input), in percentage, trained on Softmax Loss

| Class Labels | Prediction | | | |
|---|---|---|---|---|
| | Neutral | Happiness | Sadness | Anger |
| Neutral | **89.1** | 5.4 | 3.7 | 1.8 |
| Happiness | 70.5 | **25.2** | 4.3 | 0.0 |
| Sadness | 24.6 | 2.7 | **70.5** | 2.2 |
| Anger | 48.7 | 5.6 | 2.9 | **42.8** |

Table 3. Confusion matrix (ResNet on MFCC input), in percentage, trained on Focal Loss

| Class Labels | Prediction | | | |
|---|---|---|---|---|
| | Neutral | Happiness | Sadness | Anger |
| Neutral | **80.2** | 3.2 | 15.2 | 1.4 |
| Happiness | 34.7 | **52.0** | 8.5 | 4.8 |
| Sadness | 16.4 | 1.2 | **81.6** | 0.8 |
| Anger | 39.4 | 5.4 | 2.3 | **52.9** |

Additionally, we have experimented with different settings of the loss functions (softmax or focal loss) and presented the comparative study of the same in Table 4. For both inputs, spectrogram and MFCC, accuracy have increased in focal loss setting as compared to softmax loss setting. Supervision of Focal Loss helps the models to focus more towards hard-examples and down-weight the loss assigned to well-classified examples during the training process. The below comparative accuracy clearly shows the advantage of using Focal loss instead of Softmax Loss

Table 4. Ablation study of the effectiveness of Focal Loss

| Input Features | Loss functions settings | Overall Accuracy | Class Accuracy |
|---|---|---|---|
| Spectrogram | Softmax Loss | 70.2 | 55.8 |
| Spectrogram | Focal Loss | **74.2** | **64.3** |
| MFCC | Softmax Loss | 70.7 | 56.9 |
| MFCC | Focal Loss | **74.6** | **66.7** |

# 6 CONCLUSIONS

In this paper, we have proposed a speech features based ResNet architecture trained under Focal Loss, which beats the previous state-of-the-art emotion recognition accuracies. Our best model (MFCC) outperforms the benchmark results by 3.4% and 2.8% for overall and class accuracies respectively. The use of Spectrograms and MFCC provides low-level features, which when combined with ResNets has allowed us to extract very deep features boosting the model performance. With the help of Focal Loss, we have significantly improved recognition of the rarer emotion classes (Anger, Sadness and Happiness) as shown in our confusion matrices. Focal loss, which to the best of our knowledge has not been used earlier in SER systems, has helped us mitigate considerable class imbalance. Focal Loss helps to scale the standard cross-entropy loss to down-weight loss corresponding to easily classifiable examples dynamically and focus more on hard examples to make the system perform better on hard examples as well. For future work, we can experiment using Focal Loss in various architectures like CNN, RNN, and Inception Networks.